\documentclass[aps,reprint,prb,superscriptaddress,footinbib,showpacs,citeautoscr
ipt,floatfix]{revtex4-1}%
\usepackage{graphicx}
\usepackage{amssymb}
\usepackage{amsfonts}
\usepackage{amsmath}
\usepackage{bm}
\usepackage[dvipsnames,usenames]{color}
\begin{document}
\title{Spin ordering in magnetic quantum dots: From core-halo to Wigner molecules }
\author{R.~Oszwa{\l}dowski}
\affiliation{Department of Physics, University at Buffalo, Buffalo, NY 14260-1500}
\author{P.~Stano}
\affiliation{
Institute of Physics, Slovak Academy of Sciences, 845 11 Bratislava, 
Slovakia and\\
Department of Physics, University of Basel, Klingelberstrasse 82, 4056 
Basel, Switzerland}
\author{A.~G.~Petukhov}
\affiliation{Department of Physics, South Dakota School of Mines and Technology, Rapid
City, SD 57701}
\author{Igor \v{Z}uti\'{c}}
\affiliation{Department of Physics, University at Buffalo, Buffalo, NY 14260-1500}
\begin{abstract}
The interplay of confinement and Coulomb interactions in quantum dots can 
lead to strongly correlated phases differing qualitatively from the  Fermi liquid
behavior. We explore  how the presence of magnetic impurities in quantum dots
can provide additional opportunities to study correlation effects and the
resulting ordering in carrier and impurity spin.  By employing exact
diagonalization we reveal that seemingly  simple two-carrier quantum dots lead
to a rich phase diagram. We propose experiments to verify our predictions, 
in particular we discuss interband optical transitions as a 
function of temperature and magnetic field.
\end{abstract}

\maketitle

With high tunability of their parameters, quantum dots (QDs) are ideal systems 
for exploring correlation 
effects~\cite{Reimann2002:RMP,Yannouleas2007:RPP,Ghosal2006:NP,HansonRMP:2007}. 
While in 3 dimensions the correlation-induced Wigner
crystal~\cite{Wigner1934:PR} is 
elusive and only expected in the limit of an extremely low 
carrier-density~\cite{Yannouleas2007:RPP, Fulde:1993}, its nanoscale analog, 
the Wigner 
 molecule (WM)~\cite{Yannouleas2007:RPP,Ghosal2006:NP,Egger1999:PRL}, was 
observed in QDs at much higher
densities~\cite{Singha2010:PRL,Ellenberger2006:PRL}. An increase in the relative
strength of Coulomb interactions qualitatively  changes the liquid-like
independent-particle picture to that of WM characterized by electron
localization and  strong 
angular order~\cite{Ghosal2006:NP,Egger1999:PRL,Singha2010:PRL}. 

We expect that doping QDs with magnetic 
impurities~\cite{Seufert2002:PRL,Beaulac2009:S,Govorov2005:PRB,Fernandez-Rossier2004:PRL}
(typically Mn) opens unexplored opportunities to study the 
 nanoscale correlations. 
Through Mn-carrier exchange interaction, the correlations can be enhanced,
imprinted on Mn-spins, and thus observed. 
Several key advances in elucidating 
correlations in non-magnetic QDs were accomplished 
in two-electron systems~\cite{Yannouleas2007:RPP,Singha2010:PRL,%
Ellenberger2006:PRL,Pfannkuche1993:PRB}. 
However, even in simple circular QDs, identifying 
WMs is complicated by  
insufficient accuracy in the treatment of 
correlations~\cite{Yannouleas2007:RPP,Pfannkuche1993:PRB} and 
artifacts of the mean-field and Hartree-Fock approaches~\cite{HF}.
To understand these systems, the exact 
diagonalization~\cite{Singha2010:PRL,Ellenberger2006:PRL,Pfannkuche1993:PRB} 
is particularly suitable,
corroborating analytical findings for two-electron correlations 
in QDs~\cite{Merkt1991:PRB,Taut1993:PRA}.  

Here, we generalize the exact approach~\cite{Peter,Baruffa2010:PRB}
to probe the charge and spin densities of carriers as well as Mn-spin ordering 
in magnetic QDs with two carriers (holes)~\cite{note3}.
To elucidate the stability of the magnetic ordering, 
we consider different Mn-doping configurations, deformation of circular QD 
confinement,  
and examine the effects of temperature $T$ and magnetic field $B$. 

\begin{figure}[htb]
\centering
\includegraphics[width=1.0\columnwidth]{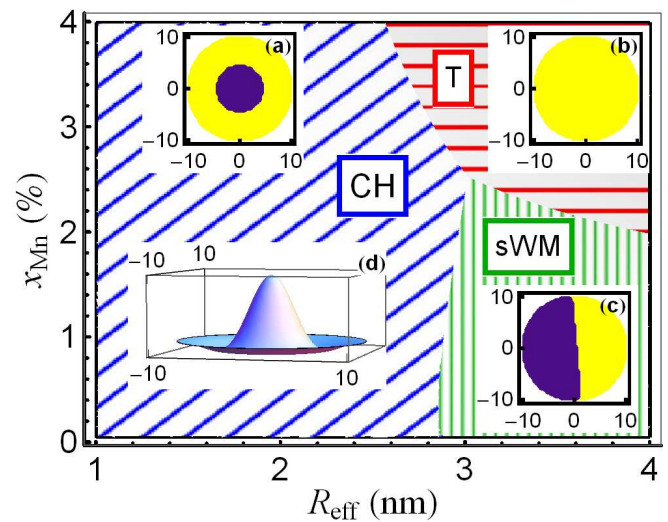}
\caption{(color online) Groundstate phase diagram
(PS vertical and skew, T horizontal hatching) as a 
function of Mn-content $x_{\rm Mn}$ and 
doping radius $R_{\rm eff}$,
for a double-occupied 
 circular dot, $\hbar \omega_{x,y}=25$ meV, at zero temperature. 
 Insets (a-c): 
QD-top view (in-plane coordinates in nm) of Mn-spin patterns; 
spins up: light, down: dark. 
(a) Core-halo (CH): $R_{\rm eff}=2$ nm,
$x_{\rm Mn}=2\%$, (b) triplet (T): $R_{\rm eff}=3.5$ nm, $x_{\rm Mn}=3\%$, 
(c) spin-Wigner Molecule (sWM): $R_{\rm eff}=3.5$ nm, $x_{\rm Mn}=1\%$, 
(d) Hole-spin density for the pattern  in (a)~\cite{APS}. 
}
\label{Fig.stab_over}
\end{figure}

The phase diagram in Fig.~\ref{Fig.stab_over}, for Mn-doped circular QDs, 
shows three magnetic groundstates, to be contrasted with the 
spin-singlet groundstate of non-magnetic QDs~\cite{Yannouleas2007:RPP}. 
The groundstate phase changes with the doping radius $R_{\rm eff}$     
and the fraction, $x_{\rm Mn}$, of cations replaced by Mn-atoms in the QD.
The carrier spin density is imprinted on Mn-spins 
forming three patterns corresponding to:  two
pseudo-singlets (PS) characterized by total hole-spin zero, 
but non-zero hole-spin density~\cite{Oszwaldowski2011:PRL}, 
and a spin triplet (T), see
 Fig.~\ref{Fig.stab_over}(a-c). 

Before we provide 
a detailed analysis, it is instructive to view the PS state in Fig.~\ref{Fig.stab_over}(c) 
as the spin-WM.  In non-magnetic circular QDs, their 
 WM ``dimer-like'' 
phase~\cite{Singha2010:PRL} can be fully revealed 
only in the pair-correlation function~\cite{Taut1993:PRA}.  
However, a similar phase can be directly detected in magnetic QDs: the 
Mn-pattern of spin-WM 
reflects a double-peaked 
hole-spin density. The 
separation between carriers, characteristic for WMs~\cite{Yannouleas2007:RPP}, is 
enhanced with Mn, which provides a spin structure. 

We use the total QD 
Hamiltonian,  $\hat{H}=\hat{H}_f+\hat{H}_{\mathrm{ex}}$, with typical
2D 
non-magnetic 
and exchange parts~\cite{Oszwaldowski2011:PRL}, where 
\begin{equation}
\hat{H}_f=\sum_{i=1,2} \left[ 
\frac{\bm{\pi}_i^2 }{2m^{\ast}}
+ \frac{m^{\ast}}{2}\left(  \omega_{x}^{2}
x_i^{2}+\omega_{y}^{2}y_i^{2}\right) \right]
+\frac{e^{2}/4\pi\epsilon} { |\bm{r}_{1}-\bm{r}_{2}|} 
\label{Eq.H},
\end{equation}
the holes are labeled by $i$, 
$m^{\ast}$ is the effective mass, $e$ the electron charge,
and $\epsilon$ 
 the dielectric constant, momentum $\bm{\pi}_i$ includes the vector potential
of field $B\!\!\parallel\!\!z$~\cite{note7}. 
The $p-d$ exchange interaction between Mn-spins and confined holes 
has the Ising form~\cite{Dorozhkin2003:PRB} 
\begin{equation}
\hat{H}_{\mathrm{ex}}=-(\beta/3)\sum_{i=1,2} \sum_{j=1}^N 
\hat{s}_{iz} \hat{S}_{jz}\ \delta\!\left(\mathbf{r}_i-\mathbf{R}_j\right),
\label{Eq.Hexch}
\end{equation} 
because of the strong $z$-axis anisotropy, arising from 
spin-orbit interaction in the 2D QDs. Here,  
$\beta$ is the exchange constant, $\hat{s}_{z}$ 
is the heavy-hole pseudospin with projections $s_{z}=\pm 3/2$, while 
$\hat{S}_z$ are operators of $z$-projection of Mn-spins positioned at 
${\mathbf R}_j$, and $N$ is the number of Mn-spins in the dot.

Since $\hat{H}_{\mathrm{ex}}$ does not contain spin-flip processes, 
the total wavefunction is a product of the hole and Mn-spin parts. 
The partition function of the system can be calculated using Gibbs
canonical distribution, 
$Z=\textrm{Tr}_{S_{jz}}\sum_n\exp\left[-E_n\left(\{S_{jz}\}\right)/k_BT\right],$
where $k_B$ is the Boltzmann constant, and
the hole eigenvalues, $E_n$, 
depend on $N$ numbers $S_{jz}$, with each $S_{jz}=-S,...,S=5/2$
(index $n$ runs over hole states for fixed $\{S_{jz}\}$).
To calculate $Z$, one would need to solve $6^N$  
replicas of the hole Schr\"odinger equation, with $N\sim 10^2 - 10^3$. 

We can overcome this obstacle of computational complexity by 
partitioning the dot into $N_c$  square cells, each containing few Mn-spins, 
and neglecting spatial variation of the 
two-hole wavefunction $\Phi_n$ within each cell through the use of 
the average hole-spin density 
\begin{equation}
\langle s_{k}\rangle_n=\frac{1}{h_z N_k}\sum_{j\in 
N_k}\left\langle\Phi_n\left|\sum_i 
\hat{s}_{iz}\,\delta(\mathbf{r}_i-\mathbf{R}_j)\right|\Phi_n\right\rangle,
\label{eq:holespin}
\end{equation} %
where $h_z$ is the QD height.
For a given cell with $N_k$ spins $S_{jz}^{(k)}$ creating a magnetic moment 
$M_k$, a distribution function of the average dimensionless magnetization, 
$m_k\equiv -M_k/g_{\rm Mn}\mu_B N_k$, 
can be expressed as
$Y(m_k)
\propto\exp\left[-G_k(m_k/S)/k_BT\right]$.
Here $g_{\rm Mn}=2$ is the Mn $g$-factor. 
The Gibbs free energy of the
$N_k$ non-interacting spins,  $G_k(m_k/S)$, is obtained by 
Legendre transformation~\cite{Petukhov2007:PRL},
\begin{equation}
\frac{G_k(x)}{N_k k_BT}\!=\!\!\left[xB_S^{-1}\!(x)\!-\!\ln\frac{\sinh\left[(1\!+\!1/2S)B
_
S^{-1}\!(x)\right]}{\sinh\left[(1/2S)B_S^{-1}\!(x)\right]}\!\right], 
\end{equation}
where $B_S^{-1}$  is the inverse of the Brillouin function $B_S$.
Using $Y(m_k)$, we 
transform $Z$  with exponential accuracy as 
\begin{equation}
\label{Eq.Zcont}
Z\propto\sum_n\int\ldots\int\exp\left[-G^n_{
\rm tot
}/k_BT\right]\prod_{k=1}^{N_c}
dm_k,
\end{equation}
where $G^{n}_{
\rm tot
}=\sum_k G_k\left(m_k/S\right)+E_n\left(\{m_k\}\right)$.
For any $n$, 
each integral in Eq.~\eqref{Eq.Zcont}  can be evaluated 
using the steepest descent method. 
Equation for the saddle point, combined with 
the Hellmann-Feynman theorem,   
$-\beta N_k\langle s_k \rangle/3 +g_{\rm Mn} N_k\mu_BB=\partial 
E_n(m_1,...,m_k,...,m_{N_c})/\partial m_k$,
leads to the self-consistency condition
\begin{equation}
m_k=SB_{S}\left[ S
\left( \beta\left \langle s_{k}\right\rangle_n/3 - g_{\rm Mn}\mu_B B \right) 
/k_B T\right]. 
\label{Eq.Brill}
\end{equation}
Our analysis shows that the self-consistency condition Eq.~(\ref{Eq.Brill})
depends on the  quantum-mechanical
average,  $\langle s_{k}\rangle_n$, 
relevant  for small systems such as QDs, rather than on the thermal 
average~\cite{Petukhov2007:PRL}, thereby avoiding artifacts arising 
from imposing the thermodynamic limit on a nanoscale system.
We use Eq.~\eqref{Eq.Brill} to find a global minimum of  
$G^0_{\rm tot}(\{m_k\})$ corresponding to the groundstate,
where $0$ stands for PS or T 
in different regions of the phase diagram. 

We find the eigenstates of $\hat{H}$
self-consistently: For fixed values of $m_k$ (randomly initialized),
we obtain intermediate two-hole states. 
Since $[\hat{H}, \hat{\Sigma}_z]$=0, where $\hat{\Sigma}_z$ is the total 
hole-spin $z$-projection, the states are either in PS
($\Sigma_z=0$) or T ($\Sigma_z=\pm 3$) orthogonal
 subspaces~\cite{Oszwaldowski2011:PRL}.
We choose the lowest state in each subspace,
use Eq.~(\ref{Eq.Brill}) to obtain new $m_k$, and restart exact diagonalization, 
iterating until convergence.
The groundstate nature, either T or PS,
depends on the QD parameters (for $x_{\rm Mn}=0=B$
the groundstate is a singlet).
 
We use ZnTe parameters: 
hole mass $m^{\ast}=0.2$ 
electron mass~\cite{Oka1981:PRB}, $\epsilon=9.4\epsilon_{0}$~\cite{Wagner1992:JCG},
$\epsilon_0$ the vacuum permittivity, $N_{0}\beta=-1.05$ eV~\cite{Sellers2010:PRB}, and $N_0=4/a^3$ 
the cation density with the lattice constant $a\simeq 6.1$ \AA.
For typical self-assembled QDs: $\hbar\omega_{x,y}= 10-30$ meV,
$x_{\mathrm{Mn}} \leq 5\%$, our standard values are $x_{\rm Mn}=1\%$, $h_z=1.8$ nm,
$\hbar\omega_0\equiv\hbar\omega_{x,y}= 25$ meV, 
corresponding to a characteristic length~\cite{Wigner} 
$l_0=(\hbar/m^{\ast}\omega_0)^{1/2}=3.9$ nm. 

We now revisit Mn-patterns in Fig.~\ref{Fig.stab_over}, calculated
at  $T=0$, with a doping profile 
$x_{\rm Mn} \{1 + \exp\left[\left(r- R_{\rm eff} \right)/ 
\xi\right] \}^{-1}$ of width $\xi= 0.25$ nm, and radius $R_{\rm eff}$,
here $r$ is the distance from the center. 
In addition to the triplet groundstate 
(all Mn-spins parallel)
at relatively large $R_{\rm eff}$ and $x_{\rm Mn}$, 
and the spin-WM groundstate at 
smaller $x_\text{Mn}$, 
another PS forms, Fig.~\ref{Fig.stab_over}(a), we term it ``core-halo" ~\cite{note4}. 
Inset (d) shows the resulting average spin-density $\left< s_k \right>$.
In contrast to spin-WM, the Mn-pattern and $\left< s_k \right>$ for core-halo 
preserve the circular symmetry of the QD. 
For  $\xi\to 0$ (no Mn for $r > R_{\rm eff}$) and 
small enough $R_{\rm eff}$ (e.g.~$R_{\rm eff}=2$ nm, $x_{\rm Mn}=1\%$) 
the core-halo patterns become purely ``ferromagnetic.'' 
We find that such unidirectional patterns 
form also for other inhomogeneous $x_\text{Mn}$-profiles,  
confirming predictions from Ref.~\cite{Oszwaldowski2011:PRL}.

\begin{figure}[htb]
\centering 
\includegraphics[width=1\columnwidth]{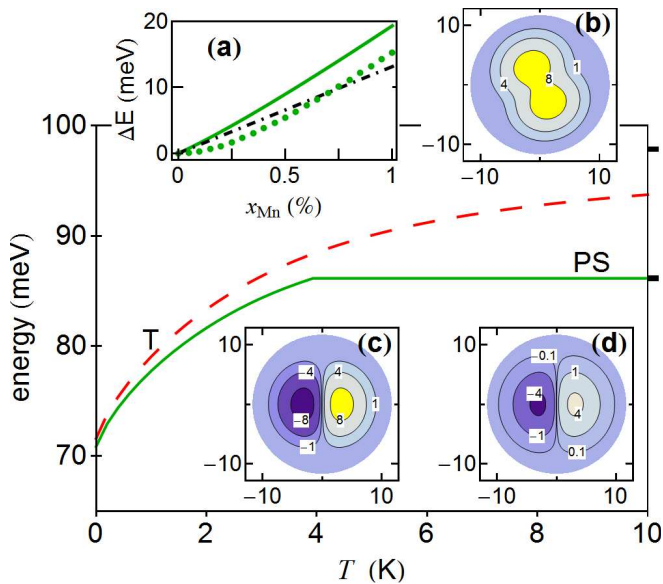}
\caption{Temperature evolution of pseudosinglet $E_{\rm PS}$ (green) and 
triplet $E_{\rm T}$ (red) energies for a 
circular QD, $\hbar \omega_{x,y}=25$ meV, and $x_{\rm Mn}=1\%$.
Bold tics show the high-$T$ (zero $p-d$ exchange) limit.
(a): Mn-induced energy gains at $T=0$ K. Dots/solid line: 
numerical/variational results~\cite{EPAPS:PRL:2012} for PS, 
$\Delta E_{\rm PS}=E_{\rm PS}\!\left(x_{\rm Mn}=0\right)-E_{\rm PS}$.
Dash-dotted line: single magnetic polaron $\Delta E_{\rm MP}=x_{\rm 
Mn}N_{0}\left|\beta\right| S/2$.  
$\Delta E_{\rm MP}<\Delta E_{\rm PS}$ for $x_{\rm Mn} \ge 0.7\%$.
Insets (b-d): $m_k$ in units of 1/4
 as a  function of position 
(nm); (b) triplet at 1 K, (c,d) PS  at 1~K, 3~K.}
\label{Fig.comp}
\end{figure}

In the $R_{\rm eff} \to \infty$ regime, we study stability of 
T and PS, termed magnetic bipolarons~\cite{Oszwaldowski2011:PRL}. 
Figure~\ref{Fig.comp}(a) shows that, already at modest 
$x_{\mathrm{Mn}}$, the Mn-induced energy gain of PS
exceeds that of  a magnetic polaron  
forming for a single carrier~\cite{Yakovlev:2010}.
This suggests that the robustness of magnetic bipolarons is similar to 
that of the well-established single magnetic polarons. 

Figure~\ref{Fig.comp} shows the temperature dependence of bipolaron energies. 
Owing to the stronger exchange interaction, the triplet approaches the high-$T$ 
limit at a higher $T$ than PS. 
The asymptotic trend $-1/T$ in $E_{\mathrm T}$, typical for paramagnets, is 
expected
since the effective exchange field  
of triplet holes, acting on Mn, is  nearly  
independent of the Mn-spin alignment. In contrast, PS  has a second-order 
transition to singlet~\cite{EPAPS:PRL:2012}.
Insets (b-d) show the magnetization  [Eq.~(\ref{Eq.Brill})] of PS and T .
Finite temperature has a
different effect on the two states.
For PS,  $m_{k}$ retains its overall 0 K shape, and decreases uniformly. 
In contrast, the saturated Mn-pattern of T, 
Fig.~\ref{Fig.stab_over}(b), undergoes a transition to a state 
with spontaneously broken circular symmetry:  two symmetrical and unidirectional
peaks appear, 
Fig.~\ref{Fig.comp}(b). The peaks reflect the hole-spin density, which
maximizes the Mn-induced energy gain through a linear combination of the two 
triplets with opposite angular momenta~\cite{Govorov2005:PRB}.

Having established the presence of magnetic bipolarons and 
Mn-patterns in circular QDs, 
it is crucial to examine the more realistic case of asymmetric confinement.
We introduce in-plane asymmetry 
through $\omega_{x,y}=\hbar /(m^{\ast} l_{x,y}^2)$, 
where $l_{x,y}= l_0 \left(1+d\right)^{\pm1}$~\cite{Elipse}.
\begin{figure}[tb]
\centering
\includegraphics[width=\columnwidth]{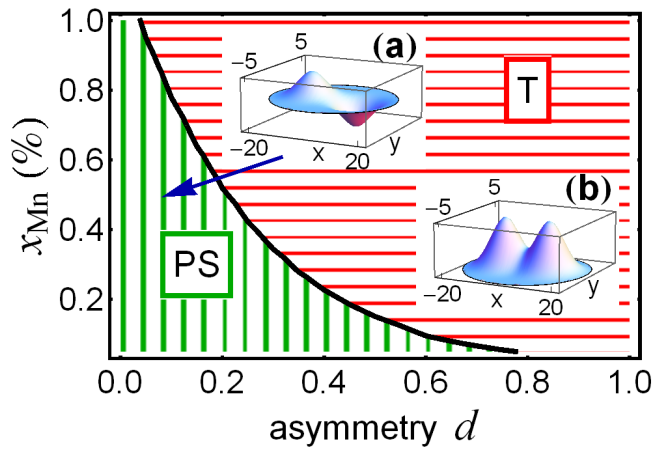}
\caption{(color online) 
Groundstate phase diagram 
(PS vertical, T horizontal 
hatching) as a function of confinement asymmetry $d$ and $x_{\rm Mn}$, at $T=0$.
Insets: hole spin density of PS (a) and T (b), $\hbar \omega_0=25$ meV, $d=1$, and 
$x_{\rm Mn}=1\%$.}
\label{Fig.stab2}
\end{figure}
The asymmetry
stabilizes the spin-WM along the weaker-potential axis, Fig.~\ref{Fig.stab2}(a), 
since separating the holes in the direction of the ``softer" potential
costs less potential energy.
As in 
circular QDs [Fig.~\ref{Fig.stab_over}, large $R_{\rm eff}$],
the triplet becomes the groundstate with increasing $x_{\rm Mn}$. 
For increasing asymmetry $d$, the 
non-magnetic singlet-triplet gap decreases, so that T becomes the groundstate
at lower $x_{\rm Mn}$ \cite{notenew}.

The results so far can be understood introducing an effective, spin-dependent 
interaction between the two holes. Each hole tends  
to polarize Mn-spins within an area defined by  
temperature and the confinement. 
Consequences of the overlap of the areas 
depend on relative hole spins. If they are 
antiparallel (PS), mutual repulsion arises~\cite{EPAPS:PRL:2012}, which would 
lead to complete separation of holes in absence of the confinement (magnetic bipolaron
would not form). 
In contrast, holes with parallel spins (T) effectively attract each 
other~\cite{Goupalov1996:SSC}, as they benefit from sharing the cloud of polarized 
Mn-spins each of them carries~\cite{EPAPS:PRL:2012}. 

Opposite limits of the doping radius $R_{\rm eff}$ offer a simple insight as to 
which of the two PS patterns is the groundstate.
For $R_{\rm eff} \to 0$, the $p-d$ exchange energy gain for spin-WM is zero, 
since this state produces $\left<s_k\right>=0$ at the QD center. 
On the other hand, for $R_{\rm eff}\to \infty$ and for weak 
confinement, the proper limit is that of two separate, localized magnetic 
polarons, a scenario consistent with spin-WM rather than 
core-halo~\cite{EPAPS:PRL:2012}.

To corroborate our predictions, as well as other works considering 
magnetic interactions in closed-shell QDs%
~\cite{Oszwaldowski2011:PRL,Fernandez-Rossier2004:PRL,Milanovic2009:PRB,%
Abolfath2012:PRL}, 
we propose experiments 
that test the existence of magnetic bipolarons,
and discriminate the different Mn-patterns.
One such probe is interband photoluminescence~\cite{Seufert2002:PRL,Beaulac2009:S,Katayama2012:JPCM}.  
With a sufficiently intense excitation, 
two kinds of emission lines appear, corresponding 
to $2 \to 1$ and $1 \to 0$ QD-occupancy transitions.
We calculate the photon energies, $E_{\rm ph}$, for the standard parameters, 
assuming:
(i) type-II band alignment~\cite{Sellers2010:PRB,EPAPS:PRL:2012},
(ii) Mn-spin pattern does not change during a recombination 
event, (iii) the system recombines from its (two- or single-hole) groundstate. 
We show that $E_{\rm ph}$ dependence on 
$T$ and $B$ allows to identify the bipolarons.
 
\begin{figure}[h]
\centering
\includegraphics[width=\columnwidth]{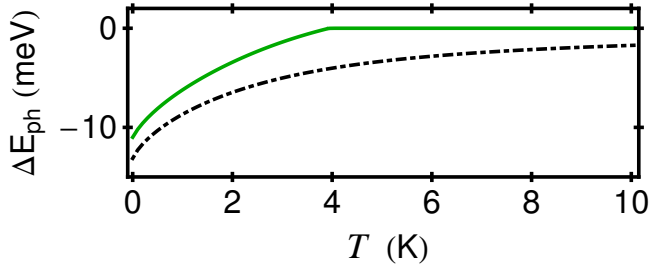}
\caption{(color online) 
Temperature  dependence of the PS $\to 1$ hole (solid), and  $1\to 0$
(dash-dotted) transition energies $E_{\rm ph}$, for  $B=0$. To better compare the 
dependencies, each line is shifted: 
$\Delta E_{\rm ph}\equiv E_{\rm ph}(T)-E_{\rm ph}(T\to \infty)$.}
\label{Fig.exp_T}
\end{figure}

We first consider varying $T$, Fig.~(\ref{Fig.exp_T}). 
Single polaron, characterized by a unidirectional Mn-pattern, shows a  
$1/T$ redshift. In contrast, thermal disruption of PS occurs with an abrupt 
change of the slope of $E_{\rm ph}$ at few kelvin.

\begin{figure}[t]
\centering
\includegraphics[width=\columnwidth]{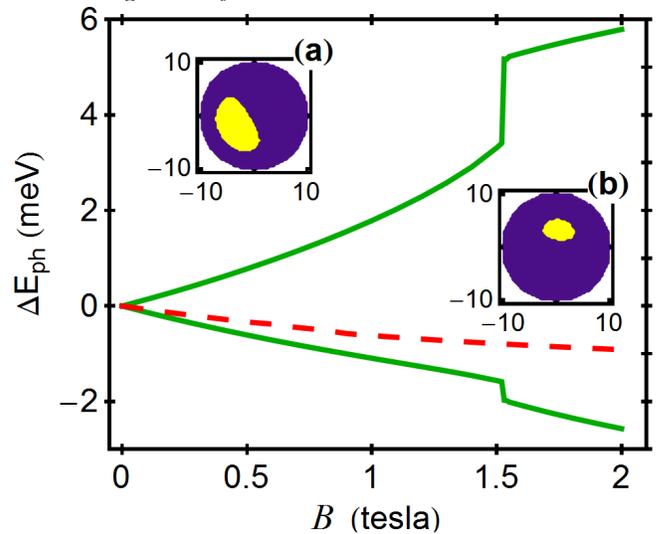}
\caption{(color online) 
$B\!\parallel\!z$  dependence of the PS $\to 1$ (solid lines) 
and T $\to 1$ (dashed), for $T=1$ K. To avoid PS$-$T crossing, we set $x_{\rm 
Mn}=0.5\%$.
Each line is shifted: 
$\Delta E_{\rm ph}\equiv E_{\rm ph}(B)-E_{\rm ph}(B=0)$.
Insets: PS Mn$-$patterns for  0.4 (a), and 1.4 tesla (b); 
light (dark) 
for $m_k>0$ ($m_k<0$).
}
\label{Fig.Bfield}
\end{figure}

We next consider 
$B\!\parallel\!z$ (Faraday 
configuration~\cite{Sellers2010:PRB,Zutic2004:RMP}).
For a triplet, $m_k\!\parallel\!-B$ everywhere. 
The small change of $\Delta E_{\rm ph}$, (Fig.~\ref{Fig.Bfield}, dashed) is 
mainly 
due to orbital effects, since $m_k$, highly saturated at the low-$T$, is 
not very sensitive to $B$. 
For PS, a part of Mn-spins 
is aligned unfavorably, i.e., parallel to $B$. Increasing $B$ changes their 
projection [Fig.~\ref{Fig.Bfield}(a,b)]. The accompanying change 
of $E_{\rm PS}$ (not shown) becomes abrupt at a threshold $B_0$ (the Mn-pattern
becomes unidirectional at $B_0$), and then 
flattens out close to its asymptotic value: the
non-magnetic singlet energy \cite{note8}. 
For $B>B_0$ ($B_0\simeq~1.5$ T in Fig.~\ref{Fig.Bfield}), the $B$-dependence of 
PS $\to 1$ transition energies
is dominated by changes of final-state energies, which react to increasing
saturation of $m_k$.
The line is split by the
field, as the remaining hole can end up  in two opposite spin states. No such 
splitting occurs for the T $\to1$ and $1\to0$ transitions. 
Thus, the splitting, and the abrupt `melting' at $B_0$, signal the PS 
groundstate.
 
Finally, the two possible PS states 
can be resolved using selection rules for 
photoluminescence~\cite{Cusack1997:PRB}. 
The circularly symmetric core-halo pattern 
forbids recombination with  $p$-like excited electron states.
However, such transitions are (weakly) allowed for 
spin-WM~\cite{EPAPS:PRL:2012}, and would 
appear at a strong optical pumping~\cite{Siebert2009:PRB}.
As an alternative to photoluminescence,
scanning tips with NV-centers
could offer  
sufficient spatial sensitivity to probe the  Mn-spin patterns~\cite{PC}. 
 
We expect that our findings will motivate future efforts to probe 
magnetization patterns and 
correlation effects at the nanoscale. Previously unexplored regimes are 
afforded with 
Mn-doping. Effective internal magnetic fields in colloidal QDs can reach 
$ \sim 100$ T~\cite{Beaulac2009:S},  beyond what is feasible with applied 
static magnetic fields. Consistent with recent advances in the field of 
nanomagnetism~\cite{Bader2006:RMP}, an increasing number of experimental 
probes are likely to meet the challenge of detecting the predicted spin ordering
even in single QDs.  

Nuclear spins in few-electron III-V QDs could provide a magnetically-active feedback
 similar to that studied here.
While the electron-nuclear spin interaction 
is weak, leading to a much smaller temperature scale for analogous polaron objects, 
this scale is known to be very strongly enhanced by electron-electron 
interactions in low-dimensional systems~\cite{Braunecker2009:PRB}. 

This work was supported by DOE-BES, ONR, 
meta-QUTE ITMS NFP 26240120022, CE SAS QUTE, EU 
Project Q-essence, APVV-0646-10 and SCIEX.

\renewcommand{\figurename}{Figure A\!\!}
\renewcommand\bibnumfmt[1]{A#1.}

\def\hksqrt{\mathpalette\DHLhksqrt}
\def\DHLhksqrt#1#2{\setbox0=\hbox{$#1\sqrt{#2\,}$}\dimen0=\ht0
\advance\dimen0-0.2\ht0
\setbox2=\hbox{\vrule height\ht0 depth -\dimen0}%
{\box0\lower0.4pt\box2}}

\renewcommand\thesection{A\arabic{section}} 
\renewcommand\thesubsection{A\arabic{section}.\arabic{subsection}} 
\setcounter{figure}{0}

\newpage

\onecolumngrid


\begin{center}

\large{\bf Spin Ordering in Magnetic Quantum 
Dots: From Core-halo to Wigner Molecules\\
(Auxiliary Material)}

\vspace{0.5cm}

\end{center}

This material provides details helping to explain some statements in 
the main text. The 1st occurrence of the EPAPS reference in 
the main text points to Sec.~\ref{Sec.VarCalc}, the 2nd and 4th to Sec.~A1.1,
 the 3rd to Sec.~A1.2, the 5th and 6th to Sec.~\ref{Sec.typeII}.

\vspace{1cm}

\twocolumngrid


\section{Variational calculation\label{Sec.VarCalc}}

We build the pseudo-singlet (PS) for the spin-WM ground state as%
\begin{align}
\Phi_\textrm{PS}=\frac{1}{\sqrt{2}}&\left[  \varphi_u\!\left(  
\bm{r}_{1}\right)
\varphi_d\!\left(  \bm{r}_{2}\right)  \chi_{\uparrow}\left(  1\right)  
\otimes
\chi_{\downarrow}\left(  2\right) \right. \nonumber\\
&\left.
 -\varphi_u\!\left(  \bm{r}_{2}\right)
\varphi_d\!\left(  \bm{r}_{1}\right)  \chi_{\uparrow}\left(  2\right)  
\otimes
\chi_{\downarrow}\left(  1\right)  \right], 
\tag{A1} \label{Eq.py}%
\end{align}
from normalized orbitals
\begin{align}
\varphi_{u}\left( \bm{r}\right) & =\frac{1}{L}\sqrt{\frac{2}{\pi 
}}\exp %
\left[ - \frac{( x-X_0 )^2+y^2 }{L^2} \right] ,
\tag{A2}\label{Eq.WR} \\
\varphi_{d}\left( \bm{r}\right) & =\frac{1}{L}\sqrt{\frac{2}{\pi 
}}\exp %
\left[ - \frac{( x+X_0 )^2+y^2 }{L^2} \right] .
\tag{A3}\label{Eq.WL}
\end{align}%
Here $\chi_{\sigma}(1)$ and $\chi_{\sigma}(2)$ are spinors
of the carriers 1 and 2, and $\sigma=\uparrow,\downarrow$, while 
$\bm{r}=(x,y)$. 
There are two variational parameters: $L$ and $X_{0}$. 
The former, width $L$, is introduced to lower the variational energy.
The role of the latter, displacement $X_0$, is more important: the exchange 
interaction 
between holes and Mn spins takes place only for $X_0\neq 0$. This parameter 
introduces separation of the holes with opposite spins, and is important for 
building the intuitive picture of effective interaction discussed in 
the main text and below.
However, the circular symmetry can be broken in any direction,
for example, the $x$-axis can be the division line between Mn spins up and 
down, this would be reflected by the following replacement
$$( x\pm X_0 )^2+y^2 \to x^2+(y\pm Y_0)^2$$
in Eqs.~(\ref{Eq.WR},\ref{Eq.WL}).

The energy without the $p-d$ exchange is%
\begin{align}
E_{\mathrm{nonm}}\left( L,X_{0}\right)  &=\frac{2\hbar ^{2}}{m^{\ast }L^{2}}%
+\frac{1}{2}m^{\ast }\omega _{0}^{2}\left( 
L^{2}+2X_{0}^{2}\right)
\tag{A4}\label{Eq.nonm} \\
+&\frac{L_{1}}{L}\sqrt{\pi ~Ry^*~\hbar \omega _{0}}\exp \left(
-2\frac{X_{0}^{2}}{L^{2}}\right) I_{0}
\left( 2\frac{X_{0}^{2}}{L^{2}}\right) ,  \notag \\
\text{where }L_1 &=\hksqrt{2}l_0=\sqrt{\frac{2\hbar }{m^\ast \omega _{0}}}.  
\tag{A5}\label{Eq.defL1}
\end{align}%
Here, $m^*$ is the effective mass, and $\hbar\omega_0$ is the energy 
quantum of the 2D harmonic oscillator.
The second line of Eq.~(\ref{Eq.nonm}) introduces the Coulomb 
repulsion of holes, 
integrated using 2D Fourier transforms 
[\citetext{A\!\!\citenum{Szabobook}}], 
$I_{0}$ is the modified Bessel function of the 1st kind, 
$Ry^{\ast}=m^\ast e^{4}/[32\left(  \pi\epsilon
 \hbar\right)  ^{2}]$ is the effective Rydberg, $e$ is the electron charge
and  $\epsilon$ is the dielectric constant.
We calculate the exchange energy using 
\begin{equation}
E_{\mathrm{ex}}=-\Delta _{0}\int \left\vert \rho_{\mathrm{s}}\left( \bm{r%
}\right) \right\vert d^{2}r, \tag{A6} \label{Eq.ExchEn}
\end{equation}%
(Eq.~(A7) in Ref.~[\citetext{A\!\!\!\citenum{EPAPS:PRL:2011}}]). Here, 
$\Delta _{0}=x_{\mathrm{Mn}}N_{0}\left\vert \beta\right\vert 5/2$
is the exchange splitting, expressed through the fraction of cations 
replaced by Mn atoms, $x_{\rm Mn}$, and the exchange constant $N_0 
\beta<0$. The 2D spin density, $\rho_{\mathrm{s}}$, is expressed in $%
\hbar $ units. It is related to  $\left<s_k\right>$ in the main text 
through $\overline{\rho}_{\mathrm 
s}\!\left(\bm{r}_k\right)/h_z=\left<s_k\right>/3$, 
where bar denotes spatial averaging over a cell $N_k$, and $h_z$ is 
the QD height. In the case of PS, the hole spin density is 
\begin{align}
\rho_{\mathrm{PS}}=\left[ \varphi_{u}^{2}\left( \bm{r}\right) 
-\varphi_{d}^{2}\left( \bm{r}\right) \right] /2. \tag{A7} 
\label{Eq.rhoPS}
\end{align}%
One finds $\int \rho_{\mathrm{PS}}d^{2}r=0$ owing to normalization of 
$\varphi_{u,d}\left( \bm{r}\right)$. We obtain the exchange energy%
\begin{equation}
E_{\mathrm{ex}}\left( L,X_{0}\right) =-\Delta _{0}\rm{erf}\left( \frac{%
\hksqrt{2}\left\vert X_{0}\right\vert }{L}\right) , \notag
\end{equation}%
where $\rm{erf}$\ is the error function. The results in Fig.~(2a main
text) are obtained by minimizing 
\begin{equation}
E_{\mathrm{tot}}=E_{\mathrm{nonm}}\left(
L,X_{0}\right) +E_{\mathrm{ex}}\left( L,X_{0}\right) \tag{A8} \label{Eq.Etot}
\end{equation}%
with respect to $%
X_{0}$ and $L$.

For $x_{\rm Mn}=0$, and without Colomb interaction, the wavefunction
in Eq.~(\ref{Eq.py}) with $X_0=0$ is the exact singlet groundstate. 
The wavefunction can be generalized to the case of 
elliptical-confinement ($\omega_x< \omega_y$) by replacing $L$ with 
the lengths $L_{x,y}=\hksqrt{2}l_{x,y}$, the latter are defined in the 
main text. The energy of the singlet is $\hbar \omega_x +\hbar \omega_y$.

The variational wavefunction for the triplet in a circular QD is the 
same as in Eq.~(4) in 
Ref.~[\citetext{A\!\!\citenum{AOszwaldowski2011:PRL}}].
The energy for the lowest triplet (T) in an elliptical QD with $x_{\rm Mn}=0$ 
is obtained by placing one of the holes
in the lowest excited orbital: $p_x$. The energy is then 
$E_{\rm T}=2\hbar \omega_x +\hbar \omega_y$. 
Thus, the singlet-triplet gap decreases as 
$\hbar \omega_x=\hbar \omega_0 \left(1+d\right)^{-2}$ with 
increasing $d$. 
The smaller this gap, the lower $x_{\rm Mn}$ value is sufficient for
the triplet to become the groundstate ($E_{\rm ex}$ for the triplet is
always more negative than $E_{\rm ex}$ for PS). This fact can be used 
to understand the dependence of the PS-T boundary on increasing $d$, 
Fig.~(3 main text).

Spin and charge densities for the triplet are proportional. Hence, the 
double-peaked spin density in  Fig.~(3b main text) can be explained in 
the variational model as a direct consequence of the form of triplet 
wavefunction: One hole is in the $s$ orbital, the other in the $p_x$ 
orbital.
\subsection{Effective repulsion for Pseudosinglet \label{Subsec.EffPS}}
It is instructive to fix $L$ and consider the trends of particular
contributions to $E_{\mathrm{tot}}$ [Eq.~(\ref{Eq.Etot})] as a function of 
$X_{0}.$ 
Kinetic 
energy does not depend on $X_{0},$ 
$E_{\mathrm{kin}}=2\hbar ^{2}/\left(m^{\ast}L^{2}\right)$.
Both Coulomb repulsion and Mn-exchange terms
decrease with increasing $X_{0}$, to zero and $-\Delta _{0}$, respectively.
The latter limiting value can be understood as the energy of two independent
magnetic polarons with opposite spin projections (hole spin-up and Mn 
spins-down for $x>0$, while hole spin down and Mn spins up for
 $x<0$). This can be seen from $\rho_{\mathrm{PS}}\left( \bm{r}\right)$ in 
Eq.~(\ref{Eq.rhoPS}), which 
has a minimum and a maximum separated, in this limit, by $2X_{0},$ see also 
Fig.~(3a main text).  Confinement (potential) energy increases as 
$m^{\ast}\omega _{0}^{2}X_{0}^{2}$,
preventing the two holes from the complete separation, so that $X_0<+\infty$,
and $\left\vert E_{\mathrm{ex}}\left( L,X_{0}<\infty \right)
\right\vert <\Delta _{0}$. Thus, we can treat $-\partial E_{\mathrm{ex}%
}\left( L,X_{0}\right) /\partial X_{0}\propto \pm \Delta_0 \exp \left[
-2X_{0}^{2}/L^{2}\right] /L$ as an effective repulsive force between
the two holes. This force enhances the Coulomb repulsion.

The above reasoning remains qualitatively correct at non-zero
temperatures, as long as they are are lower than the phase transition
temperature. For $T>0$,
we calculate the variational exchange energy from a generalization of 
Eq.~(\ref{Eq.ExchEn})%
\begin{equation}
E_{\mathrm{ex}}=-\Delta _{0}\int \rho_{\mathrm{s}}\left( \bm{r}\right)
B_S\left( \left\vert \beta \right\vert \frac{5}{2}\frac{\rho_{\mathrm{s}%
}\left( \bm{r}\right) }{h_zk_{B}T}\right) d^{2}r, \tag{A9} 
\label{Eq.ExchEnT}
\end{equation}%
resulting from Eqs.~(2, 6 main text).
Eq.~(\ref{Eq.ExchEn}) above is obtained in the limit of $T\rightarrow 0$, when 
the Brillouin function 
$B_s\left[ \left\vert \beta \right\vert 5\rho_{\mathrm{s}%
}\left( \bm{r}\right) /\left(2 h_z k_{B}T\right) \right] \rightarrow 
\mathrm{%
sign}\left[ \rho_{\mathrm{s}}\left( \bm{r}\right) \right]$. 

\begin{figure}
\centering
\includegraphics[width=1.0\columnwidth]{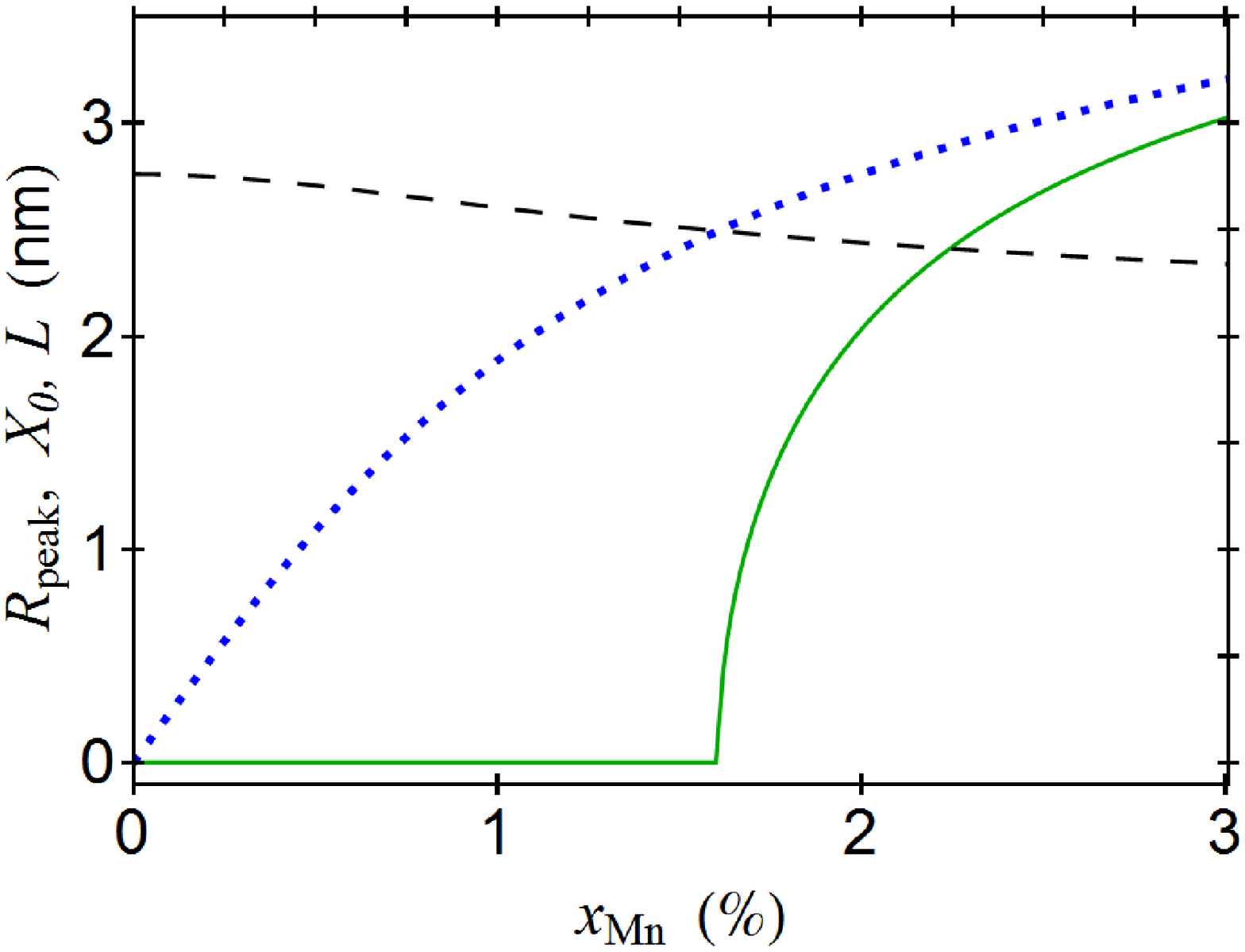}%
\caption{(color online) Variational results for positions of maxima of charge 
density 
$\varrho$ 
for PS at $T=0$. 
Solid green line: $R_{\rm peak}$, dotted blue line: $X_0$, dashed black line: 
$L/2$.
We use the standard QD parameters defined in 
the main text, except for the effective Rydberg $Ry^*=0$.}
\label{Fig.eff_attr_intro}
\end{figure}

Before numerically exemplifying the above model of effective interaction, we 
notice that the numerical hole-spin density $\left< s_k \right>$ would not be a 
convenient choice for studying the 
effective repulsion in PS, as it vanishes in the small $x_{\rm Mn}$ limit. 
However, the spatial separation of the two holes can also be seen in the charge 
density (in units of $|e|$)
$$\varrho \left( \bm{r}\right) =
\varphi_u\left( \bm{r}\right)^2+\varphi_d\left( \bm{r}\right)^2.$$
The distance of any of the two symmetrical maxima from the QD center, $r=0$,
is given by
$$R_{\rm peak}=X_0\tanh \left[ \frac{4R_{\rm peak}X_0}{L^2}\right]$$
in the variational framework.
To satisfy this condition for $R_{\rm peak}>0$, the condition $X_0>L/2$ must 
hold. We neglect Coulomb interaction for simplicity for the rest of this 
subsection. For $x_{\rm Mn}\rightarrow 0$ we find $L\to L_{1}$, 
[Eq.~(\ref{Eq.defL1})], and $X_{0}\rightarrow 0,$ so there is a lower-limit
$x_{\rm Mn}$, below which 
$\varrho$ 
has only one extremum; a maximum 
at $r=0$. Above this $x_{\rm Mn}$, there are two maxima at 
$\bm{r}=\left(\pm R_{\rm peak},0\right)$ and a minimum at $r=0$. 
The results of these considerations are shown in Fig.~A\ref{Fig.eff_attr_intro}.

\subsection{Effective attraction for Triplet \label{Subsec.EffTr}}
In the case of the triplets discussed in the main text,
the hole spin density $\rho _{\mathrm{T}}$ has a constant sign, 
so that $\int \rho_{\mathrm{T}}d^{2}r=\pm 1$. Thus, at zero temperature,
the maximum energy gain
occurs for unidirectional alignment of Mn spins, and no 
changes of the wavefunction are needed to attain the maximum,
cf.~Eq.~(\ref{Eq.ExchEn}). 
This suggests that the $T=0$ limit is not helpful in the case of the triplets, 
since at any $T>0$, the local exchange energy, $\left\vert \beta \right\vert 
\rho_{\mathrm{s}}\left( \bm{r}\right) /h_z$, 
is smaller than $k_{B}T$ for Mn spins sufficiently remote from the
center, so that some localization (change of wavefunction) is 
beneficial. As expected, the magnetization patterns for the triplets remain 
unidirectional at any temperature. 

\subsection{Effective interaction in the Exact Diagonalization method}

\begin{figure}
\centering
\includegraphics[width=1.0\columnwidth]{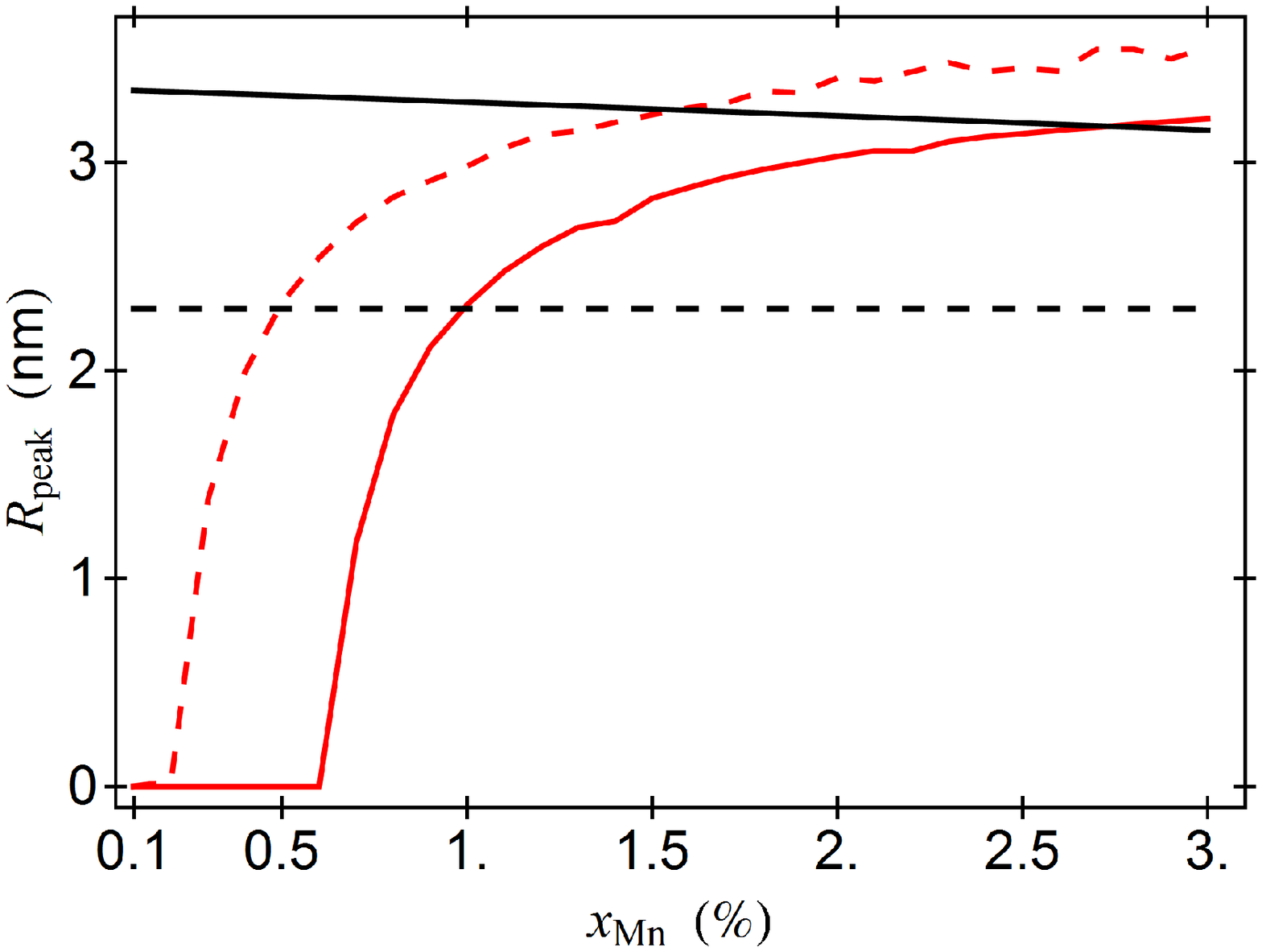}%
\caption{(color online) Distance, $R_{\rm peak}$, from the center of a QD to 
the position
of one of two symmetrical maxima in the charge density 
$\varrho$,
calculated in 
EXD. $x_{\rm Mn}>0.1\%$ as explained in the text here. Red: PS, black: 
triplet. Dashed: $T=0$, solid $T=2$~K.
The oscillations at high $x_{\rm Mn}$ are artefacts due to numerical precision.
We use the standard QD parameters defined in the main text, taking into account
the Coulomb interaction.
}
\label{Fig.eff_attr}
\end{figure}

Figure~A\ref{Fig.eff_attr} shows the numerical results obtained by the exact 
diagonalization (EXD) method described in the main text. 
 The smallest Mn-content used is $x_{\rm Mn}=0.1\% $. For this value, the 
number of Mn spins in the disk-shape volume of radius $L_1$ and height $h_z$ is 
$N=3$. Hence, for the very low 
contents $x_{\rm Mn}\le 0.1\% $, the continuum approximation would no 
longer be justifiable. Typical values of $x_{\rm Mn}$ in experiments are 
higher than 0.1\%.

Same as in the variational approach, we find that in EXD the distance 
$R_{\rm peak}$ depends on $x_{\rm Mn}$, justifying the notion of effective 
interaction. Note the opposite behavior of PS vs.~triplet at $T>0$, 
representing 
repulsion and attraction respectively, as discussed in
the main text.


Comparing the EXD results for PS with the variational ones 
(Fig.~A\ref{Fig.eff_attr_intro}), we see that the off-center maxima start to 
appear at a much lower value of $x_{\rm Mn}$ (0.6 vs.~1.6\%). This is because 
the Coulomb interaction, 
present in EXD, favors separation of holes. 

Finally, we note that a different quantum approach to interaction of magnetic 
bipolarons was used in Ref.~[\citetext{A\!\citenum{Bednarski2012:JoPCM}}].
\subsection{Core-halo pattern}
As explained in the main text, a different Mn-spin pattern forms for some
Mn-doping profiles. This pattern can be studied variationally by replacing the 
$%
\varphi_{u}$ and $\varphi_{d}$ functions in Eq.~(\ref{Eq.py}) by two
Gaussians with $X_{0}=0,$ but with different widths: $L_{u}$ for spin up and 
$L_{d}$ for spin down, see Eq.~(3) in 
Ref.~[\citetext{A\!\citenum{AOszwaldowski2011:PRL}}].

\section{Photoluminescence in type-II QD\lowercase{s}\label{Sec.typeII}}

\begin{figure}[htb]
\centering
\includegraphics[width=0.7\columnwidth]{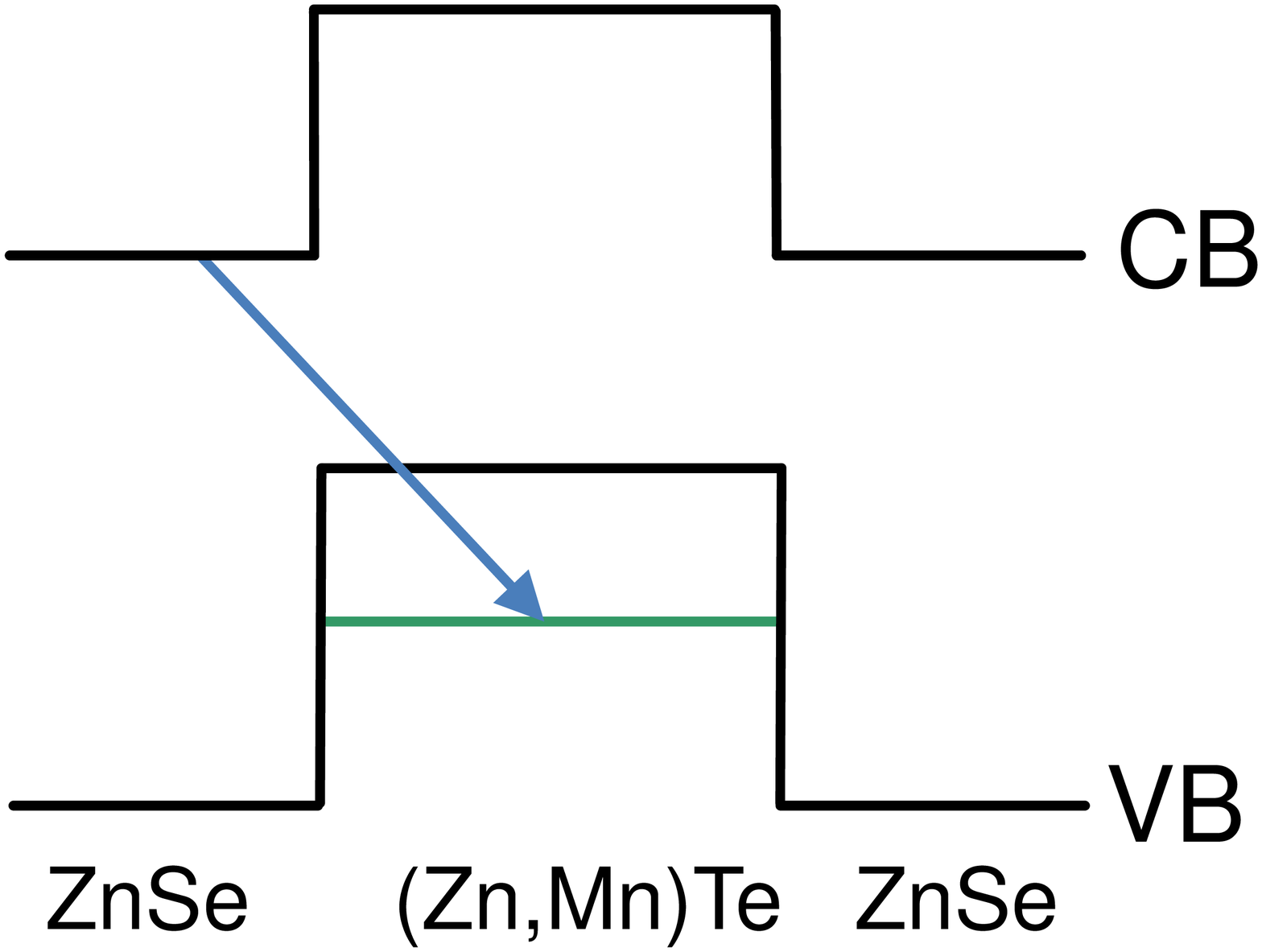}%
\caption{(color online) Band alignment for a type-II band profile, found
in, e.g., ZnSe/ZnTe QDs. CB and VB stand for conduction and valence bands 
respectively.
}
\label{Fig.typeII}
\end{figure}

In a (Zn,Mn)Te/ZnSe QD system, holes are confined in the QDs, while
electrons are in the barrier (weakly bound to holes through
Coulomb interaction) [\citetext{A\citenum{ASellers2010:PRB,Kusk2007}}].
This is an example of the type-II band profile (Fig.~A\ref{Fig.typeII}),
as opposed to the type-I profile, where both electrons and holes are confined.
Thus, the exchange interaction of Mn with electrons is much weaker than with
holes. 

\begin{figure}
\centering
\includegraphics[width=1.0\columnwidth]{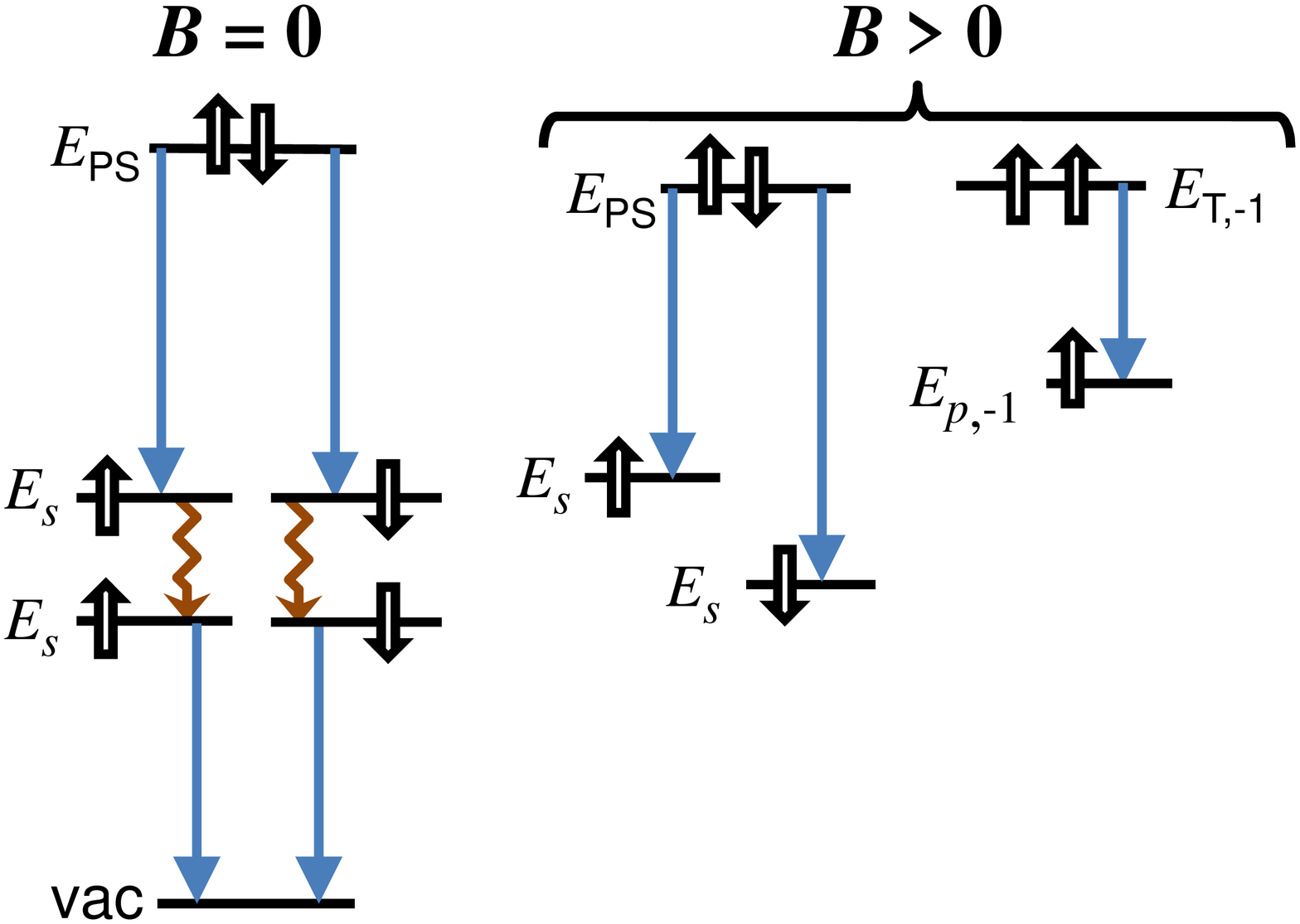}%
\caption{(color online) Jab{\l}o\'nski diagram 
[\citetext{A\citenum{Lichtman2005:NM}}] of 
the transitions discussed in the main text.
Horizontal black lines: single and two-hole levels, as well as the zero-hole
groundstate, ``vacuum". Blue arrows: radiative transitions. 
Zig-zag arrows: non-radiative relaxation of Mn spins.
Hollow arrows: spins of holes. An electron spin (not shown) must be antiparallel
to hole spin to recombine [\citetext{A\citenum{AZutic2004:RMP}}].
The energies of the radiative transitions in the left/right panel are shown in 
Figs.~4/5 of main text.
$E_{\rm PS}$/$E_{\rm T}$ is the pseudo-singlet/triplet energy. $E_s$ or $E_p$ 
are the energies of a single hole in an $s$-like or $p$-like state respectively.
The number $-1$ is the angular momentum of the triplet and $p$-like 
states.
}
\label{Fig.spectra}
\end{figure}

Figure ~A\ref{Fig.spectra} shows the transitions calculated in the manuscript.
We assume that owing to fast energy relaxation, luminescence from groundstates 
dominates over other initial states. Two types of transitions occur, in which 
the QD occupancies change as: $2\to 1$ and $1\to 0$.
These two kinds of lines are separated by an energy 
determined mainly by hole repulsion (charging energy) $\sim 35$ meV.
In general, the final state of a $2\to 1$ is not the initial
state of the subsequent $1\to 0$ transition, since relaxation of a 
Mn-spin pattern may occur between the two radiative events.
The numerical calculations of the final states of the $2\to 1$ transitions are 
not 
selfconsistent, but rather carried out for Mn-magnetization $m_k$ fixed in
the pattern established when two holes are still present.

As argued in the main text, interband transitions from $p$-like states of 
electrons may arise due to breaking of the circular-symmetry by the Mn-pattern 
(the non-magnetic confinement potential remaining circularly symmetrical). The 
symmetry 
selection rules for the interband transitions are given by overlaps of electron 
and hole wavefunctions. Expanding the shifted Gaussian orbitals, 
Eqs.~(\ref{Eq.WR},\ref{Eq.WL}) for small $X_0/L$, we obtain
\begin{align}
\varphi_{u}\left( \bm{r}\right) & \simeq s\left( \bm{r}\right) 
+\frac{X_{0}}{L}p_{x}\left( \bm{r}\right),\notag \\
\varphi_{d}\left( \bm{r}\right) & \simeq s\left( \bm{r}\right) 
-\frac{X_{0}}{L}p_{x}\left( \bm{r}\right),\notag
\end{align}%
where $s=\sqrt{2/\pi}\exp \left( -r^{2}/L^{2}\right)/L$, and
$p_{x}=\hksqrt{8/\pi}~x\exp \left( -r^{2}/L^{2}\right)/L^2$.
Thus, for $x_{\rm Mn}>0$, the hole wavefunction contains $p_x$ 
orbitals, which have the same symmetry as electron $p$-like orbitals.
The hole wavefunction for the core-halo pattern does not have the $p$
orbitals in the equivalent expansion.
\begin{acknowledgments}
We thank Jeongsu Lee and Karel Vyborny for help in calculations for 
this material.
\end{acknowledgments}

\begin{thebibliography}{99}
\bibitem{note7}
The action of $B$ on magnetic moments of holes is negligible for the effects 
described here.%
\bibitem{note4}
The actual orientation of spins in core-halo and spin-WM patterns, as well as the 
azimuthal angle of the division line in the latter, are not determined by $\hat{H}$,
but pseudorandomly in the numerical procedure.
\bibitem{notenew} 
The phase boundaries (Figs.~\ref{Fig.stab_over} and \ref{Fig.stab2}) 
result from interplay of the
single-particle confinement, Coulomb interaction, and magnetic exchange. 
The phase boundaries will shift in function of the QD parameters.
\bibitem{note8} For $T=0$ (full saturation), $E_{\rm PS}$, as well as the 
final-state (single-hole) energies are constant above $B_0$.
\end{thebibliography}

\end{document}